\documentclass[onecolumn, showpacs, preprintnumbers,
nofootinbib, aps, prd, longbibliography, superscriptaddress,
10pt, showkeys, notitlepage, eqsecnum]{revtex4-1}

\usepackage{graphicx,amssymb,amsmath,amsthm,amsfonts,epsfig,times}
\usepackage[linktocpage]{hyperref}
\usepackage[usenames,dvipsnames]{color}
\usepackage{epstopdf}
\usepackage{xcolor}
\usepackage{url}

\usepackage{cleveref}
\usepackage{times}

\newcommand{\tn}{\textnormal}

\definecolor{darkmidnightblue}{rgb}{0.0, 0.2, 0.4}
\definecolor{venetianred}{rgb}{0.78, 0.03, 0.08}
\definecolor{indigo}{rgb}{0.29, 0.0, 0.51}
\definecolor{jazzberryjam}{rgb}{0.65, 0.04, 0.37}
\definecolor{pinegreen}{rgb}{0.0, 0.47, 0.44}
\definecolor{coolblack}{rgb}{0.0, 0.18, 0.39}

\usepackage{tikz}
\usetikzlibrary{shapes.geometric, arrows}

\tikzstyle{startstop} = [rectangle, rounded corners, minimum width=3cm, minimum height=1cm,text centered, draw=black, fill=orange!15] 
\tikzstyle{paper} = [rectangle, minimum width=3cm, minimum height=1cm,text centered, draw=black, fill=blue!5] 
\tikzstyle{textbox} = [rectangle, rounded corners, minimum width=3cm, minimum height=1cm,text centered, draw=black, text width=4cm, align=center, fill=orange!20] 
\tikzstyle{process} = [rectangle, minimum width=3cm, minimum height=1cm, text centered, draw=black, fill=orange!30] 
\tikzstyle{invisible} = [rectangle, rounded corners, minimum width=3cm, minimum height=1cm, text centered, draw=yellow, fill=white] 
\tikzstyle{arrow} = [,->,>=latex] 
\tikzstyle{arrowd} = [dashed,->,>=latex] 

\hypersetup{colorlinks=true, citecolor=coolblack, linkcolor=coolblack,
urlcolor = coolblack}

\begin{document}
\title{Compact objects in Horndeski gravity}

\author{Hector O. Silva}
\email{hosilva@phy.olemiss.edu}
\affiliation{Department of Physics and Astronomy, The University of Mississippi, University, Mississippi 38677, USA}

\author{Andrea Maselli}
\email{andrea.maselli@roma1.infn.it}
\affiliation{Theoretical Astrophysics, IAAT, Eberhard-Karls University of
T\"ubingen, T\"ubingen 72076, Germany}

\author{Masato Minamitsuji}
\email{masato.minamitsuji@ist.utl.pt}
\affiliation{Departamento de F\'isica, CENTRA, Instituto Superior
T\'ecnico, Universidade de Lisboa, Avenida Rovisco Pais 1,
1049 Lisboa, Portugal}

\author{Emanuele Berti}
\email{eberti@olemiss.edu}
\affiliation{Department of Physics and Astronomy, The University of Mississippi, University, Mississippi 38677, USA}
\affiliation{Departamento de F\'isica, CENTRA, Instituto Superior
T\'ecnico, Universidade de Lisboa, Avenida Rovisco Pais 1,
1049 Lisboa, Portugal}

\begin{abstract}
  Horndeski gravity holds a special position as the most general
  extension of Einstein's theory of general relativity with a single
  scalar degree of freedom and second-order field equations. Because
  of these features, Horndeski gravity is an attractive
  phenomenological playground to investigate the consequences of
  modifications of general relativity in cosmology and
  astrophysics. We present a review of the progress made so far in the
  study of compact objects (black holes and neutron stars) within
  Horndeski gravity.  In particular, we review our recent work on
  slowly rotating black holes and present some new results on slowly
  rotating neutron stars.
\end{abstract}

\maketitle

\tableofcontents

\section{Introduction}
\label{sec:intro}

Einstein's theory of general relativity (GR) has passed all
experimental tests in its centennial history with flying
colors~\cite{Will:2014kxa}. Most precision tests of GR in our Solar
System are confined to the weak-field/slow-motion regime. An exception
are binary pulsars, where the orbital motion is nonrelativistic but
the individual objects have strong gravitational
fields~\cite{Wex:2014nva}. As we witness the birth of the era of
gravitational-wave astronomy~\cite{Abbott:2016blz}, in the coming
years we can hope to test GR in its strong field regime -- as in the
recent detection of binary black hole (BH)
mergers~\cite{TheLIGOScientific:2016src}, or possibly in the future
via neutron star (NS) mergers -- and in its radiative regime, e.g. by
searching for the additional polarizations modes of gravitational
radiation predicted by competing theories. Observational and
theoretical issues with Einstein's theory -- such as the unknown
nature of dark matter and dark energy, the presence of curvature
singularities and the search for an ultraviolet completion of GR --
have motivated strong efforts to develop modified theories of gravity
which differ from GR in the infrared and ultraviolet regimes, while
being consistent with the stringent observational bounds at
intermediate energies~\cite{Berti:2015itd}.  Testing GR and searching
for signatures of any deviation from its predictions is a major goal
of several areas of research, including
cosmology~\cite{Clifton:2011jh}, ``standard'' electromagnetic
astronomy~\cite{Psaltis:2008bb,Bambi:2015kza}, and Earth- and
space-based gravitational-wave
astronomy~\cite{Yunes:2013dva,Gair:2012nm}.

In this work we will consider a very general modification of GR known
as Horndeski gravity. The theory has its origins in the 1970s with
Horndeski's attempt~\cite{Horndeski:1974wa} to obtain the most general
action for a scalar-tensor theory with a single scalar degree of
freedom and second-order field equations. Horndeski gravity has
attracted much interest recently. One motivation has been the study of
scalar-tensor theories with self-tuning
cosmologies~\cite{Charmousis:2011bf,Charmousis:2011ea}. Horndeski's
theory was rediscovered in the context of Galileon theories, i.e.,
scalar-tensor models which in flat space-time have Galilean
symmetry. The generalization of Galileon theories to an arbitrary
number of dimensions~\cite{Deffayet:2009mn} was shown to be equivalent
to Horndeski gravity in four dimensions \cite{Kobayashi:2011nu}.  The
theory can also be obtained by a Kaluza-Klein compactification of
higher-dimensional Lovelock
gravity~\cite{VanAcoleyen:2011mj,Charmousis:2014mia}. Tensor-multiscalar
theories~\cite{Damour:1992we,Rainer:1996gw,Horbatsch:2015bua,Kuusk:2015dda}
and multiscalar versions of Horndeski
gravity~\cite{Deffayet:2010zh,Padilla:2012dx,Charmousis:2014zaa,Ohashi:2015fma}
have also been formulated, but they will not be our main focus here.

In this paper we review our current understanding of compact objects
(BHs and NSs) in Horndeski gravity with a single scalar field.  This
topic has received increasing attention because the study of compact
objects can allow us to better understand the theory and (potentially)
to confront it against observations in astrophysical
settings. Progress has been rapid, and a summary of the recent
developments in this field seems quite timely.

The paper is organized as follows. In Sec.~\ref{sec:horndeski} we
review the basic aspects of Horndeski gravity and discuss some special
cases. In Sec.~\ref{sec:bhsol} we review BH solutions in the theory
and their stability properties. We also review no-hair theorems, their
validity and loopholes. In Sec.~\ref{sec:nss} we discuss NSs,
presenting some new results for slowly rotating stars. In
Sec.~\ref{sec:conclusions} we point out some directions for future
research.

\section{Overview of Horndeski's theory of gravity}
\label{sec:horndeski}

We start by reviewing Horndeski gravity in its modern formulation.
The action of the theory reads
\begin{equation}
S=\sum_{i=2}^{5}\int d^{4}x\sqrt{-g}{\cal L}_i\ ,
\label{eq:action}
\end{equation}
where
\begin{subequations}
\begin{align}
{\cal L}_2&=G_2\ ,\\
{\cal L}_3&=-G_{3}\square\phi\ ,\\
{\cal L}_4&=G_{4}R+G_{4\tn{X}}\left[(\square\phi)^2-\phi_{\mu\nu}^2\right]\ ,\\
{\cal L}_5&=G_{5}G_{\mu\nu}\phi^{\mu\nu}-\frac{G_{5\tn{X}}}{6}\left[(\square\phi)^3+2\phi_{\mu\nu}^3 -3\phi_{\mu\nu}^2\square\phi\right]\ .
\label{eq:lagrangean}
\end{align}
\end{subequations}
Here $g_{\mu\nu}$ is the metric tensor and $g\equiv {\rm det}(g_{\mu\nu})$ its
determinant. The Ricci scalar and Einstein tensor associated with $g_{\mu\nu}$
are denoted by $R$ and $G_{\mu\nu}$, respectively.
The functions $G_{i}=G_{i}(\phi,X)$ depend only on the scalar
field $\phi$ and its kinetic energy $X=-\partial_\mu\phi\partial^\mu\phi/2$.
We use units such that $m_{\rm Pl}^2\equiv(8\pi G)^{-1}=1$, where $m_{\rm Pl}$
is the reduced Planck mass. For brevity we have also defined the shorthand notation
$\phi_{\mu\dots\nu}\equiv \nabla_\mu\dots\nabla_\nu\phi$,
$\phi_{\mu\nu}^2 \equiv \phi_{\mu\nu}\phi^{\mu\nu}$,
$\phi_{\mu\nu}^3 \equiv
\phi_{\mu\nu}\phi^{\nu\alpha}\phi^{\mu}{_{\alpha}}$
and $\Box\phi\equiv g^{\mu\nu} \phi_{\mu\nu}$.

An attractive feature of Horndeski gravity is its generality. The
theory includes a broad spectrum of phenomenological dark energy
models, as well as modified gravity theories with a single scalar
degree of freedom (in this review we will not discuss ``beyond
Horndeski'' theories~\cite{Zumalacarregui:2013pma,Gleyzes:2014dya} or
extensions of Horndeski including two or more scalar degrees of
freedom~\cite{Padilla:2012dx,Charmousis:2014zaa,Ohashi:2015fma}).

Some important special limits of the theory are listed below:

\begin{enumerate}
\item GR is obtained by choosing $G_4=1/2$ and $G_2=G_3=G_5=0$.
\item When the only nonzero $G_{i}$ function is $G_4=F(\phi)$, we
  recover a scalar-tensor theory with nonminimal coupling of the form
  $F(\phi)R$. Consequently, Brans-Dicke theory and $f(R)$ gravity are
  special cases of Horndeski gravity.
\item Einstein-dilaton-Gauss-Bonnet (EdGB) gravity, whose action is
  \begin{equation}\label{eq:EdGBaction}
    S=\int d^4x
\sqrt{-g}\left(\frac{1}{2}R+X+\xi(\phi) R^2_{\tn{GB}}\right)\ ,
  \end{equation}
where
$R^2_\tn{GB}=R^2-4R_{\mu\nu}R^{\mu\nu}+R_{\alpha\beta\gamma\delta}R^{\alpha\beta\gamma\delta}$
is the Gauss-Bonnet invariant, corresponds to the choices
\begin{subequations}
\begin{align}
G_2&=X+8\xi^{(4)} X^2(3-\ln X)\,,\quad
G_3=4\xi^{(3)}X(7-3\ln X)\,,\label{EDGBK}\\
G_4&=\frac{1}{2}+4\xi^{(2)}X(2-\ln X)\,, \quad
G_5=-4\xi^{(1)}\ln X\,,\label{EDGBG5}
\end{align}
\end{subequations}
where $R_{\alpha\beta\gamma\delta}$ and $R_{\mu\nu}$ are the
Riemann and Ricci tensors, and we have defined
$\xi^{(n)}\equiv \partial^n \xi/\partial \phi^n$~\cite{Kobayashi:2011nu}.
\item A theory including nonminimal derivative coupling between the
  scalar field $\phi$ and the Einstein tensor $G_{\mu\nu}$ (the
  ``John'' Lagrangian in the language of the so-called ``Fab Four''
  model~\cite{Charmousis:2011bf,Charmousis:2011ea}), with action
\begin{equation}\label{eq:nonminact}
S=\int d^{4}
x\sqrt{-g}\left[\zeta R+2\beta X+\eta G^{\mu\nu}\phi_\mu\phi_\nu
-2\Lambda_0\right]\,,
\end{equation}
can be constructed by setting
\begin{equation}
G_{2}=-2\Lambda_0+2\beta X\ , \quad
G_4=\zeta+\eta X\ , \quad
G_3=G_5=0\ ,
\end{equation}
where $\Lambda_0$, $\eta$, $\zeta$ and $\beta$ are constants. Note
that a coupling of the form $G^{\mu\nu}\phi_{\mu}\phi_{\nu}$ can also
be obtained by setting $G_5=-\phi$ and integrating by
parts~\cite{Kobayashi:2014eva}. This action also arises in the decoupling limit of massive gravity ~\cite{deRham:2011by,Heisenberg:2014kea}.
\label{itm:phieins}
\item The Lagrangian ${\cal L}_2$ corresponds to the k-essence
  field~\cite{ArmendarizPicon:2000ah,ArmendarizPicon:1999rj,Alishahiha:2004eh}.
  For this reason, in some of the literature the function $G_2$ is
  denoted by $K$.
\label{itm:kess}
\item\label{itm:covgal} The covariant Galileon
  model~\cite{Deffayet:2009wt} is recovered by setting $G_2=-c_2X$,
  $G_3=-c_3 X/M^3$, $G_4=M_{\rm Pl}^2/2-c_4X^2/M^6$ and
  $G_5=3c_5X^2/M^9$, where the $c_i$ ($i=2,\ldots,5$) are constants
  and $M$ is a constant with dimensions of mass.
\end{enumerate}

Because of the generality of Horndeski gravity, a comprehensive review
of compact objects would inevitably have to discuss important
subclasses that have been studied for a long time, such as EdGB and
$f(R)$ gravity\cite{Berti:2015itd}. For brevity we will focus on the
subclasses that have not been reviewed in the past (i.e., the special
cases \ref{itm:phieins}--\ref{itm:covgal} above). We will also focus
on four-dimensional solutions.

\section{Black hole solutions}
\label{sec:bhsol}

As mentioned in the introduction, Horndeski gravity received renewed interest
because of its applications to cosmology. Only more recently BH solutions
have been obtained and studied in several subclasses of the theory.
We begin this section by reviewing an important no-hair theorem established
by Hui and Nicolis~\cite{Hui:2012qt}, which sets tight constraints on the
search for hairy BH solutions.

\subsection{A no-hair theorem in Horndeski gravity}
\label{sec:nohair}

Hui and Nicolis~\cite{Hui:2012qt} presented a no-hair theorem which is
valid for shift-symmetric Horndeski gravity, i.e., the subclass of the
Horndeski action which remains invariant under a transformation
$\phi \rightarrow \phi + c$ of the scalar field, where $c$ is a
constant. The theorem is applicable to vacuum, static, spherically
symmetric and asymptotically flat BHs~\footnote{ A no-hair theorem for
the subclass~\ref{itm:phieins} was obtained by Germani et
al.~\cite{Germani:2011bc} following a different strategy.}.
The shift symmetry implies the existence of a conserved current $J^{\mu}$
for the scalar field which satisfies $\nabla_{\mu}J^{\mu} = 0$.

We assume a line element of the form
\begin{equation}
ds^2 = -A(r) dt^2 + B(r)^{-1} dr^2 + r^2 (d\theta^2 + \sin^2\theta\,d\varphi^2)\,.
\label{eq:lineelement}
\end{equation}
The proof of the theorem relies on the following steps/assumptions:
({\it i}) the scalar field is assumed to have the same symmetries as
the metric, implying that the only nonzero component of $J^{\mu}$ (if
any) is the radial component $J^{r}$. ({\it ii}) We require
$J^2 \equiv J^{\mu} J_{\mu} = (J^{r})^2/B$ to remain regular at the
horizon $r_{\rm h}$. Since $B(r_{\rm h}) = 0$, we must set $J^{r} = 0$
at the horizon. ({\it iii}) From $\nabla_{\mu} J^{\mu} = 0$, we obtain
$\partial_{r} J^{r} + 2 J^{r}/r = 0$, with solution $J^{r} r^2 = k$,
where $k$ is an integration constant. As $r_{\rm h} \neq 0$, condition ({\it ii})
at the horizon directly implies $k = 0$. Therefore $J^{r} = 0$ $\forall$ $r$.
({\it iv}) It is argued that $J^r$ has the schematic form
\begin{equation}
J^{r} = B \phi' F(g,g',g'',\phi')\,,
\label{eq:currentnh}
\end{equation}
where $F$ is a generic function of the metric, its derivatives
(indicated by the primes) and the derivatives of the scalar field
$\phi'$. Asymptotic flatness requires that $B \rightarrow 1$ and
$\phi' \rightarrow 0$ at spatial infinity, while $F$ tends to a
nonzero constant. The latter condition follows from the requirement
that in the weak-field limit the kinetic energy be quadratic in $\phi$
and that $J^{\mu} \approx \partial^{\mu} \phi$, up to a normalization
constant. Moving ``inwards'' towards the horizon, $\phi'$ can become
nonzero, and $B$ and $F$, by continuity, remain nonzero. We then
conclude that $J^{r} \neq 0$, contradicting the conclusion from ({\it
  iii}). This can be resolved by forcing $\phi' = 0$ $\forall$
$r$. Consequently the scalar field must be constant, and by exploiting
the shift symmetry we can set its value to zero.

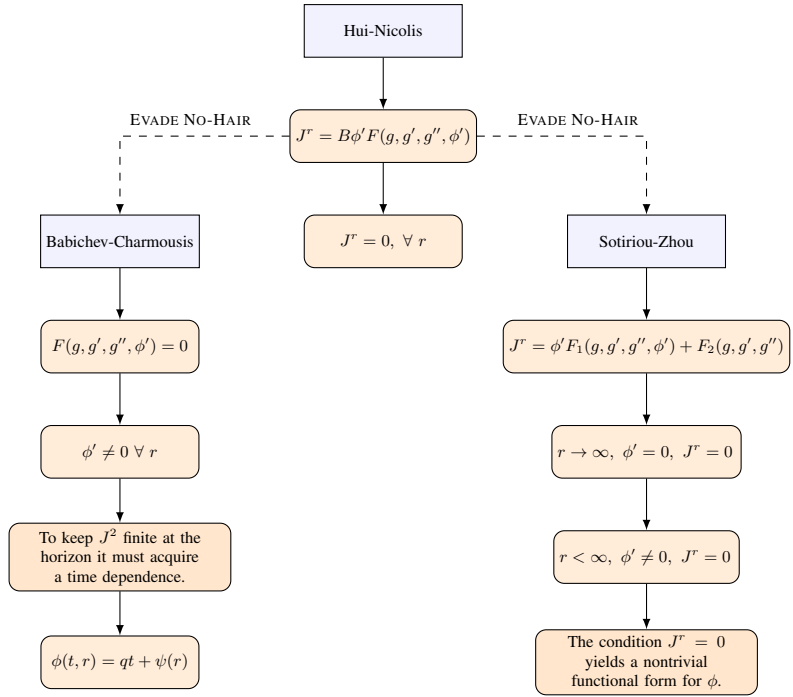
\begin{figure}
\centering
\begin{tikzpicture}[node distance=2cm, scale=0.7, transform shape]

\node (start) [paper] {Hui-Nicolis};

\node (current) [startstop, below of=start] {$J^{r} = B \phi' F(g,g',g'',\phi')$};

\node (nocurrent) [startstop, below of=current] {$J^{r} = 0,\,\, \forall \,\, r$};

\node (babchar) [paper, left of=nocurrent, xshift=-3cm] {Babichev-Charmousis};

\node (sotzhou) [paper, right of=nocurrent, xshift=3cm] {Sotiriou-Zhou};


\node (letbabchar) [startstop, below of=babchar] {$F(g,g',g'',\phi') = 0$};

\node (letbabchar2) [startstop, below of=letbabchar] {$\phi' \neq 0 \,\, \forall \,\, r$};

\node (concbabchar) [textbox, below of=letbabchar2] {To keep $J^2$ finite at the
horizon it must acquire a time dependence.};

\node (finbabchar) [startstop, below of=concbabchar] {$\phi(t,r) = q t + \psi(r)$};

\node (letsotzhou) [startstop, below of=sotzhou] {$J^{r} = \phi'F_1(g,g',g'',\phi') + F_2(g,g',g'')$};

\node (letsotzhou2) [startstop, below of=letsotzhou] {$
r \rightarrow \infty, \,\,
\phi' = 0, \,\,
J^{r} = 0\,$};

\node (letsotzhou3) [startstop, below of=letsotzhou2] {$
r < \infty, \,\,
\phi' \neq 0, \,\,
J^{r} = 0\,$};

\node (finsotzhou) [textbox, below of=letsotzhou3] {The condition $J^r = 0$ yields
a nontrivial functional form for $\phi$.};

\draw [arrow] (start) -- (current);
\draw [arrow] (current) -- (nocurrent);
\draw [arrowd] (current) -| node[above, xshift=-1.3cm, yshift=0.1cm] {{\sc Evade No-Hair}} (sotzhou);
\draw [arrow] (sotzhou) -- (letsotzhou);
\draw [arrowd] (current) -| node[above, xshift=1.3cm, yshift=0.1cm] {{ \sc Evade No-Hair}} (babchar);
\draw [arrow] (babchar) -- (letbabchar);
\draw [arrow] (letbabchar) -- (letbabchar2);
\draw [arrow] (letbabchar2) -- (concbabchar);
\draw [arrow] (concbabchar) -- (finbabchar);
\draw [arrow] (letsotzhou) -- (letsotzhou2);
\draw [arrow] (letsotzhou2) -- (letsotzhou3);
\draw [arrow] (letsotzhou3) -- (finsotzhou);

\end{tikzpicture}
\caption{A schematic representation of the Hui-Nicolis no-hair theorem for
shift-symmetric Horndeski gravity, and two possible ways of violating it.}
\label{fig:scheme}
\end{figure}

\subsection{Static hairy black hole solutions}
\label{sec:bhsolutions}

As pointed out by Sotiriou and Zhou~\cite{Sotiriou:2013qea}, the
no-hair theorem summarized above has a loophole allowing for hairy BH
solutions. Furthermore, other hairy solutions can be obtained by
relaxing some of the assumptions that enter the proof of the no hair
theorem. BH hair is classified as either primary (described by an
independent charge, e.g. a scalar charge) or secondary (depending on
other charges, such as the mass $M$ of the BH). The BH solutions known
so far in Horndeski gravity have secondary
hair~\cite{Herdeiro:2015waa,Volkov:2016ehx}.  The various
possibilities are schematically summarized in Fig.~\ref{fig:scheme},
and discussed below.

\subsubsection{Asymptotically flat black holes}
\label{sec:bhflat}

The loophole pointed out by Sotiriou and Zhou concerns the last step
of the proof.  An explicit calculation of the current $J^{\mu}$ using
the action of Eq.~(\ref{eq:action}) reveals that the form of $J^{r}$
assumed by Hui and Nicolis does not necessarily hold for all
shift-symmetric Horndeski theories. Explicitly, $J^{r}$ reads
\begin{align}
J^{r} &= -B G_{\rm{2X}}\phi' + \frac{B^2 \phi'^{2}}{2} \left(\frac{A'}{A}
+ \frac{4}{r} \right) G_{\rm{3X}} + \frac{2 B^2 \phi'}{r}
\left( \frac{A'}{A} - \frac{1}{Br} + \frac{1}{r}\right)G_{\rm{4X}} \nonumber \\
&-\frac{2 B^3 \phi'^3}{r} \left( \frac{A'}{A} + \frac{1}{r}\right) G_{\rm{4XX}}
-\frac{B^3 \phi'^{2}}{2 r^2} \frac{A'}{A}\left( \frac{3B-1}{B}\right)G_{{\rm 5X}}
+\frac{A'}{A}\frac{B^4 \phi'^4}{2 r^2} G_{{\rm 5XX}}\,,
\label{eq:completejr}
\end{align}
where we note that to impose shift symmetry in the theory we must set
$G_{i}(X,\phi) = G_{i}(X)$. Observe that all the terms involve powers
of $\phi'$, as required by Hui and Nicolis.
%
%
In particular the first term has the form (\ref{eq:currentnh}) and
$F \rightarrow - G_{\rm 2X}$ at spatial infinity, as required by the theorem.
The other terms depend on derivatives of $A$, $B$ and/or inverse powers of $r$,
and seem to vanish for large $r$, as required.
%
In principle, however, hairy BHs could exist for theories where the
functions $G_{i}(X)$ are chosen such that $J^{r}$ contains terms
independent of $\phi'$, but no negative powers of $\phi'$
(cf. Fig.~\ref{fig:scheme}). Another alternative would be to have
negative power of $\phi'$, however this generally corresponds to
theories that would not admit flat space with a trivial scalar
configuration as a solution, leading to violations of local
Lorentz symmetry~\cite{Sotiriou:2013qea}.

An explicit example~\cite{Sotiriou:2014pfa} of a theory in which
$J^{r}$ contains terms independent of $\phi'$ is EdGB theory, with a
linear coupling $\xi(\phi) = \alpha \phi$ -- where $\alpha$ is a
constant -- between the scalar field and the Gauss-Bonnet invariant in
Eq.~(\ref{eq:EdGBaction}). Exploiting shift symmetry and Lovelock's
theorem, Sotiriou and Zhou showed that, in fact, this is the unique
shift-symmetric subclass of Horndeski for which this happens. For this
theory, $J^{r}$ has the form
\begin{equation}
J^{r}_{\rm EdGB} = -B\phi' - 4\alpha \frac{A'}{A}\frac{B(B-1)}{r^2}\,.
\label{eq:curedgb}
\end{equation}
The current vanishes at infinity, however the second term allows for
scalar hair growth when $J^{r} = 0$, i.e. $\phi'$ is nontrivial. In
this theory, the hair is of the ``second kind,'' i.e. it depends on
the mass $M$ of the BH~\cite{Sotiriou:2013qea,Herdeiro:2015waa}.

For comparison, it is instructive to write down the form of $J^{r}$
for a theory with nonmininmal derivative coupling
(cf. item~\ref{itm:phieins}). The nonvanishing component of the
current in this case is
\begin{equation}
J^{r}_{\rm Gg} = B\phi' \left[ -2\beta + \frac{2B}{r}\left( \frac{A'}{A}
-\frac{1}{Br} + \frac{1}{r}\right)\eta\right]\,,
\end{equation}
which is of the form assumed by Hui and Nicolis, and therefore the
theory does not admit asymptotically flat hairy BH
solutions~\cite{Germani:2011bc}.

An alternative approach was considered by Babichev and
Charmousis~\cite{Babichev:2013cya} (and further explored
in Ref.~\cite{Charmousis:2015aya}).
For the nonminimal derivative coupling theory, the conserved
current $J^{\mu}$ associated with the shift symmetry can be written as
\begin{equation}
J^{\mu} = (\beta g^{\mu\nu} - \eta G^{\mu\nu})\partial_{\nu} \phi\,,
\end{equation}
which opens two possibilities to satisfy the condition $J^{r} = 0$:
({\it i}) set the scalar field to be constant or ({\it ii})
set $\beta g^{\mu\nu} - \eta G^{\mu\nu} = 0$, and then allow
$\partial_{\nu} \phi$ to be nonzero.

In general, this latter condition cannot be satisfied together with
the regularity of $J^2$ at the horizon. Babichev and Charmousis show,
however, that both conditions can be satisfied if the scalar field is
allowed to be time-dependent, i.e.  $\phi = \phi(t,r)$, while the
background metric is still fixed by Eq. (\ref{eq:lineelement}):
see Fig~\ref{fig:scheme}. In particular, they find hairy solutions with
\begin{equation}
\phi(t,r) = qt + \psi(r)\,.
\label{eq:babicharphi}
\end{equation}
Among the solutions constructed in this way, the non-minimally coupled
theory with $\beta = \Lambda = 0$ [cf. Eq.~(\ref{eq:nonminact})] admits
a ``stealth'' solution, where a Schwarzschild BH metric supports a
nontrivial, regular scalar field configuration which does not
backreact on the spacetime. We stress that although $\phi(t,r)$
diverges at future infinity, it is the derivatives of $\phi(t,r)$
which appear in the action, and these remain well-behaved due to the
linear dependence on $t$.

Assuming shift {\it and} reflection symmetry ($\phi \rightarrow -\phi$), which
implies that $G_3 = G_5 = 0$ in the action (\ref{eq:action}),
Kobayashi and Tanahashi~\cite{Kobayashi:2014eva} extended the
Babichev-Charmousis approach to obtain a very general class of BH
solutions which do not require specific assumptions on the form of
$G_2$ and $G_4$. A key ingredient in the derivation is that $X$ is a
constant. Their solutions are regular, in the sense that $X$ and $J^2$
are well-behaved.

In principle the time dependence of the scalar field could affect the
$tr$ component of the field equations ${\cal E}_{tr}$. However,
assuming diffeomorphism invariance, shift symmetry and that
$\phi(t,r) = q t + \psi(r)$, Babichev et al.~\cite{Babichev:2015rva}
showed that ${\cal E}_{tr}$ is proportional to $J^{r}$. As we have
seen, regularity of the current $J^{\mu}$ at the horizon demands that
$J^{r} = 0$ everywhere, and consequently ${\cal E}_{tr} = 0$.

\subsubsection{Non-asymptotically flat spacetimes}
\label{sec:bhds}

To our knowledge, Rinaldi~\cite{Rinaldi:2012vy} was the first to
explore BH solutions in the special class of Horndeski's theory given
by Eq.~(\ref{eq:nonminact}) with $\Lambda_0 = 0$.  The scalar field
was found to be imaginary because $\phi'(r)^2<0$ outside the horizon,
which may imply an instability of the solution. Note however that
because of shift symmetry the field equations of the theory never
contain $\phi$, but only its derivative; in this sense, one could
think of $\phi'(r)^2$ as a separate field. As shown by Anabalon et
al.~\cite{Anabalon:2013oea} and
Minamitsuji~\cite{Minamitsuji:2013ura}, the presence of a cosmological
constant $\Lambda_0$ cures this problem. Requiring that the scalar
field remains real imposes certain constraints on the parameters
$\Lambda_0$, $\eta$ and $\zeta$. Self-tuning BHs with de Sitter
asymptotics were also obtained~\cite{Babichev:2013cya,Charmousis:2015aya}.
A BH solution which asymptotically
approaches a Lifshitz spacetime was also found using the
Babichev-Charmousis construction, and therefore a time-dependent
scalar field~\cite{Bravo-Gaete:2013dca}.

Some works have also considered Horndeski gravity in the presence of a
Maxwell field. For instance, considering the theory with nonminimal
coupling between the scalar field and the Einstein tensor with an
additional Maxwell Lagrangian $\propto F_{\mu\nu}F^{\mu\nu}$, Cisterna
and Erices~\cite{Cisterna:2014nua} obtained electrically charged BH
solutions which are asymptotically anti-de Sitter (AdS). An
interesting solution is obtained when the scalar field dynamics is
determined solely by the ``John'' Lagrangian in
Eq.~(\ref{eq:nonminact}), i.e when $\beta = 0$.  In this case one
finds a charged BH solution which is locally flat as
$r \rightarrow \infty$, with an asymptotically constant electric field
$E \propto \Lambda_0$ supported by the presence of the cosmological
constant. Extending their previous work~\cite{Babichev:2013cya},
Babichev and Charmousis also studied a more general class of
Horndeski-Maxwell theories, allowing for all the possible couplings
between the Maxwell and scalar field under the $U(1)$ and shift
symmetries~\cite{Babichev:2015rva} and obtaining charged BH solutions.
Kolyvaris et al.~\cite{Kolyvaris:2011fk} also obtained solutions
involving scalar and Maxwell fields.

\subsection{Stationary hairy black hole solutions}
\label{sec:bhrot}

All of these works considered static, spherically symmetric
solutions. The construction of slowly rotating solutions was recently
studied in the Hartle-Thorne
formalism~\cite{Hartle:1967he,Hartle:1968si}, where rotation is
considered as a perturbation on an otherwise static spherically
symmetric background~\cite{Maselli:2015yva}. Let us consider the line
element
\begin{equation}
ds^2 = -A(r) dt^2 + B(r)^{-2} dr^2 + r^2 (d\theta^2 + \sin^2\theta\, d\varphi^2) -
2\, \omega(r) r^2 \sin^2\theta\, dt d\varphi\,,
\label{eq:ht}
\end{equation}
where the function $\omega$ is related with the dragging of inertial
frames, and it is of the same order as the angular velocity
$\Omega$. Under the slow-rotation assumption we can write down the
full set of field equations of Horndeski gravity, without assuming
shift or reflection symmetries, and for a scalar field of the form
(\ref{eq:babicharphi}). We found that for all the solutions reported
by Kobayashi and Tanahashi, $\omega$ behaves exactly as in GR, i.e.
$\omega = k_1 + k_2 /r^3$, where $k_1$ and $k_2$ are integration
constants~\cite{Maselli:2015yva}.

We also formulated an extension of the Hui-Nicolis no-hair theorem
including both a time-dependent scalar field of the form
(\ref{eq:babicharphi}) and slow rotation.  Following the arguments
outlined in Sec.~\ref{sec:nohair}, in our case $J^2$ becomes
\begin{equation}
J^2 = \frac{(J^r)^2}{B} - (J^t)^2 A\,.
\end{equation}
Imposing regularity at the horizon yields again $J^r = 0$, as long as
$J^t$ does not diverge. In general,
this imposes certain restrictions on $A$ and $B$. In the particular
case of shift- and reflection-symmetric theories, these restrictions
reduce to the condition that $(B/A)'$ remain finite at the
horizon. This time, because of the time dependence,
$\nabla_{\mu} J^{\mu} = 0$ gives
$\partial_r J^{r} + 2 J^r/r + \partial_t J^{t} = 0$. An explicit
calculation of $J^{t}$ reveals that $\partial_t J^{t} = 0$
(a particular consequence of the linear time dependence of $\phi$), and
therefore, as in the Hui-Nicolis case, we conclude that $J^r = 0$
$\forall$ $r$. As we mentioned before, although the scalar field
depends on time, $J^{r} = 0$ implies that ${\cal E}_{tr} = 0$. At
last, the current $J^r$ can still be written in the schematic form
(\ref{eq:currentnh}), and therefore we conclude that $\phi'=0$ by the
same arguments as in the Hui-Nicolis proof. This result can be used to
justify the absence of scalar field corrections to GR in our
slow-rotation approximation. As in the original theorem, our
generalized result can be violated either using the Sotiriou-Zhou
loophole (i.e., in EdGB theory) or by demanding that
$F(g,g',g'',\phi') = 0$, which imposes restrictions on the choices of
the functions $G_{i}$.

In their study of odd-parity gravitational perturbations in the
nonminimal derivative coupling theory (see Sec.~\ref{sec:stability}),
Cisterna et al.~\cite{Cisterna:2015uya} reached the same conclusion:
the frame-dragging equation (in vacuum) is the same as in GR.

In general, however, we expect that rotation will cause ``bald''
slowly rotating BH solutions to grow hair. At second perturbative
order in the slow-rotation expansion -- i.e. when we add terms of
order $\Omega^2$ to Eq.~(\ref{eq:ht}) -- the scalar field also gets
corrected~\cite{Sotiriou:2013qea}:
\begin{equation}
\phi(r,\theta) = \phi^{(0)}(r) + \phi^{(2)} (r,\theta)\,,
\end{equation}
where superscripts indicate the perturbative order. Therefore the
scalar field will not, in general, have spherical symmetry. The
$\phi^{(2)}$ correction is likely to affect the form of $J^{r}$, and
one could expect that the current will no longer have the form of
Eq.~(\ref{eq:currentnh}). Moreover, the presence of a nontrivial
component $J^{\theta}$ of the current should play a role when
demanding that $J^2$ is well-behaved at the horizon.

Two possible approaches to tackle this problem would involve
considering higher-order corrections in the Hartle-Thorne
scheme~\cite{MaselliPrep} or constructing fully numerical
solutions. Both approaches have been successfully applied to
rotating BHs in EdGB theory~\cite{Pani:2009wy,Maselli:2014fca,Maselli:2015tta,Kleihaus:2014lba,Kleihaus:2011tg},
which have nontrivial scalar hair.

In conclusion, the study of general rotating BHs in Horndeski theory
remains a fairly unexplored and interesting topic.

\subsection{Stability, quasinormal modes and collapse}
\label{sec:stability}

In a tour de force calculation, the odd~\cite{Kobayashi:2012kh} and
even~\cite{Kobayashi:2014wsa} gravitational perturbations of static,
spherically symmetric backgrounds were studied by Kobayashi and
collaborators. When applied to particular subclasses of Horndeski
gravity, their perturbation equations yield conditions preventing the
appearance of ghost and gradient instabilities. The analysis of these
papers assumes the scalar field to be static, and therefore the
results cannot be applied to the
Babichev-Charmousis~\cite{Babichev:2013cya} and
Kobayashi-Tanahashi~\cite{Kobayashi:2014eva} solutions, where the
scalar field depends linearly on time. Focusing on the shift- and
reflection-symmetric sectors of the theory, Ogawa et
al.~\cite{Ogawa:2015pea} analyzed odd gravitational perturbations
allowing the scalar field to be time-dependent. A surprising result is
that solutions with $X$ =constant~\cite{Kobayashi:2014eva} suffer
either from ghost or gradient instabilities in the vicinity of the
horizon.


Minamitsuji~\cite{Minamitsuji:2014hha} investigated the stability of
BH solutions under massless scalar perturbations in the nonminimal
derivative coupling subclass~\cite{Minamitsuji:2013ura}. The solutions
are asymptotically AdS, so the calculation can be done using the same
techniques used for Schwarzschild-AdS BHs~\cite{Horowitz:1999jd}.  The
quasinormal modes can be computed, and no unstable modes were
found. Considering the same BH solutions, Cisterna et
al.~\cite{Cisterna:2015uya} found that BHs are stable under odd-parity
gravitational perturbations (see Anabalon et
al.~\cite{Anabalon:2014lea} for an earlier study).

Let us also remark that the gravitational collapse of the scalar field
was studied by Koutsoumbas et al. in the nonminimal derivative
coupling theory~\cite{Koutsoumbas:2015ekk} .

\section{Neutron stars}
\label{sec:nss}

NSs in Horndeski gravity have received attention only very
recently. Cisterna et al.~\cite{Cisterna:2015yla} considered the
subclass of Horndeski's theory involving a nonminimal coupling between
the scalar field and the Einstein tensor. They wrote down the
generalized Tolman-Oppenheimer-Volkoff (TOV) equations and adopted the
same assumptions that Babichev and Charmousis~\cite{Babichev:2013cya}
used to obtain stealth BH solution, constructing asymptotically flat
NS models.  Numerical integration of the TOV equations revealed that,
depending on the sign of the coupling constant $\eta$ in
Eq.~(\ref{eq:nonminact}), the mass-radius relation is shifted either
upwards ($\eta < 0$) or downwards ($\eta > 0$) with respect to GR. A
second constant $q$ [cf. Eq. (\ref{eq:babicharphi})] controls
deviations from GR of the Horndeski NS model for a given value of
$\eta$. Interestingly, an expansion of the pressure equation near the
star's center allows one to constrain the allowed values of
($q,\eta$) for a given central energy density $\epsilon_{\rm c}$ by
demanding that the pressure monotonically decay within the star
(similar considerations permit to constrain the parameter space of NSs
in EdGB theory~\cite{Pani:2011xm}).

Applying the results of Maselli et al.~\cite{Maselli:2015yva} and
Cisterna et al.~\cite{Cisterna:2015uya} to the vacuum exterior region,
we find that the frame-dragging equation outside slowly rotating stars
in this subclass of Horndeski gravity is identical to the
frame-dragging equation in GR. As in the case of BHs, this is not
expected to hold at higher orders in the slow-rotation expansion. This
might have interesting consequences for the structure of NSs. For
example, the quadrupole moment could differ from GR in the presence of
a scalar field, and this may affect astrophysical observables, such as
quasi-periodic oscillations (QPOs).

One of the outstanding unresolved problem in testing modified theories
of gravity using NSs is how to unambiguously disentangle the
uncertainties in the equation of state from the effects predicted by modified
gravity~\cite{Berti:2015itd,Glampedakis:2015sua}.

The ``universal'' (nearly equation of state independent) relations
between the moment of inertia $I$, the quadrupole moment $Q$ and the
tidal Love number $\lambda$ ($I$-Love-$Q$ relations) found in
GR~\cite{Yagi:2013bca,Yagi:2013awa} can help alleviate this problem.
These $I$-Love-$Q$ relations were studied in a broad set of modified
theories of
gravity~\cite{Pani:2014jra,Kleihaus:2014lba,Yagi:2013awa,Sham:2013cya,Doneva:2014faa,Doneva:2015hsa},
and it will be interesting to see whether they hold in Horndeski
gravity.

\begin{figure}[h]
\includegraphics[width=0.467\columnwidth]{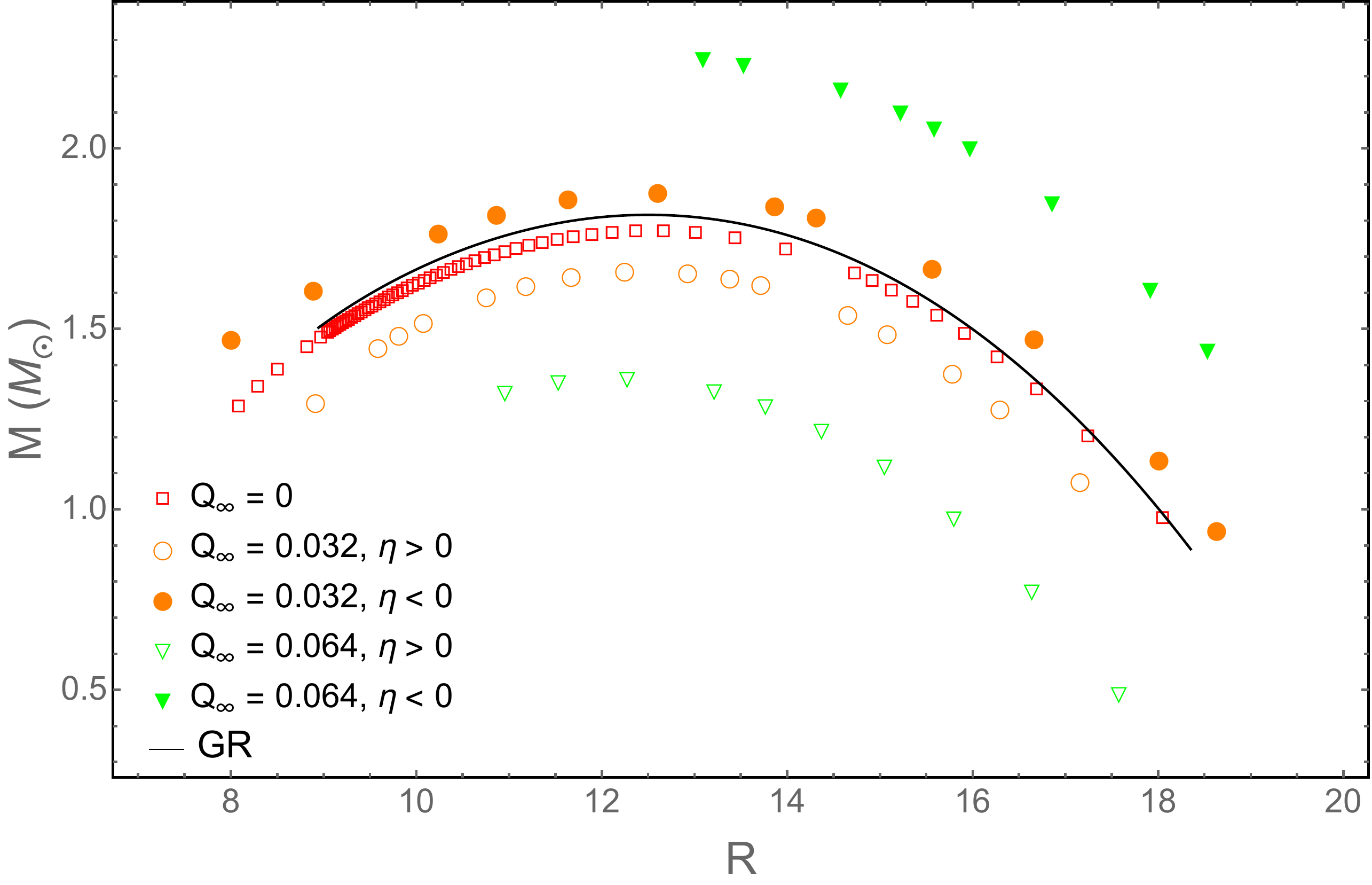}
\includegraphics[width=0.46\columnwidth, scale=0.01]{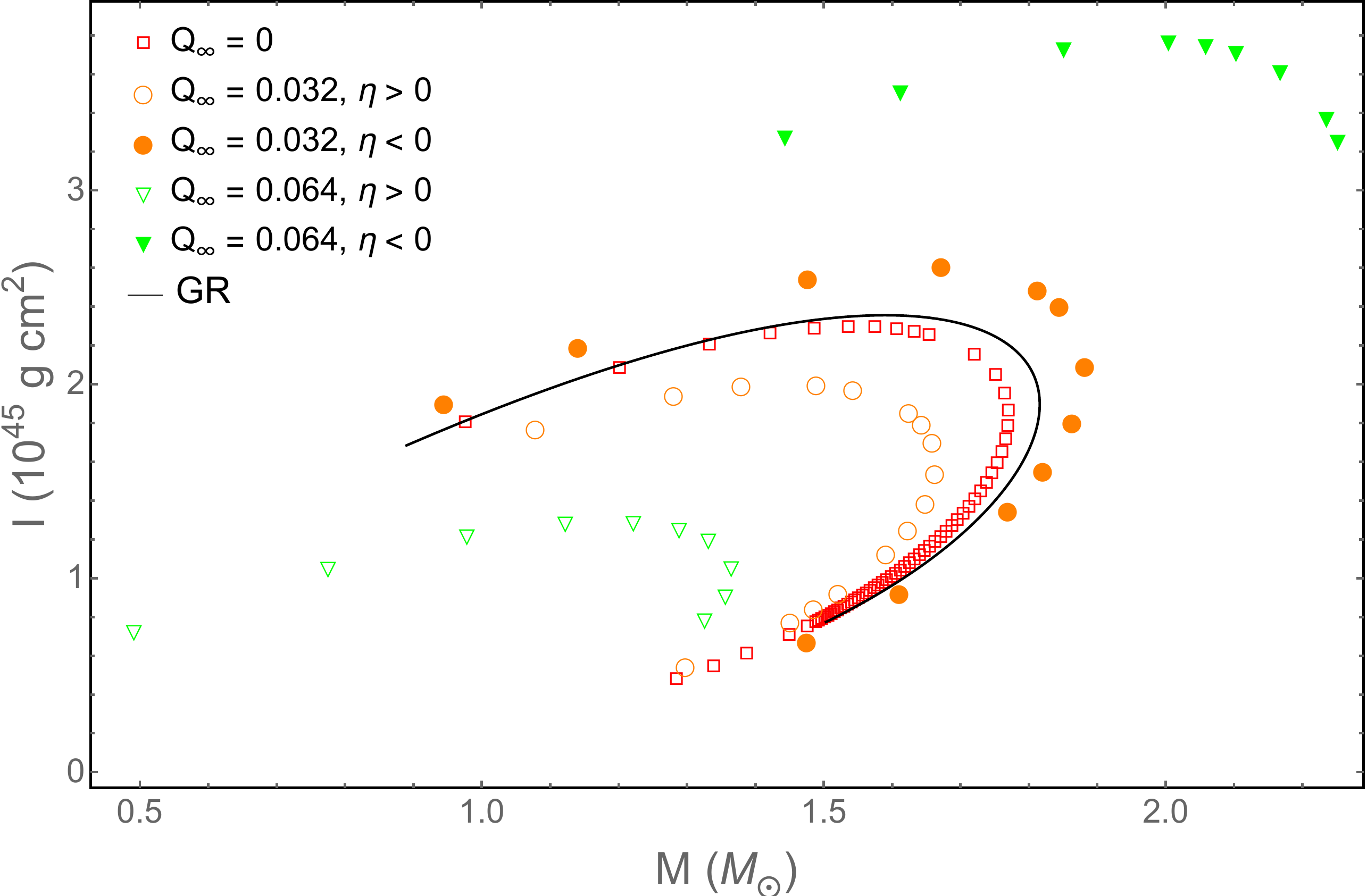}
\caption{{\it NSs in Horndeski gravity.} We show how the constants
$Q_{\infty}$ and $\eta$ affects the bulk properties of slowly rotating NSs.
{\it Left:} Mass versus radius. {\it Right:} Moment of inertia versus mass.
The solid curves represent the general relativistic solutions. The constant
$Q_{\infty}$ is related with $q$: cf. Cisterna et al.~\cite{Cisterna:2015yla}
for details.}
\label{fig:srhorn}
\end{figure}

In Fig.~\ref{fig:srhorn} we show some preliminary results in this
direction. We compute the mass, radius and moment of inertia of NSs in
Horndeski gravity by numerically integrating the same stellar
structure equations as in Cisterna et al.~\cite{Cisterna:2015uya},
generalized to include the effect of rotation at first order in the
Hartle-Thorne perturbative scheme (which allows us to compute the
moment of inertia $I$).  For illustrative purposes, here we consider a
polytropic equation of state. The effect of the unconstrained parameters $\eta$ and
$Q_{\infty}$ on the bulk properties of the star can be large.  We will
present a more detailed and extensive study (including NS models in
other subclasses of Horndeski gravity) in a forthcoming
paper~\cite{MaselliPrep}.

Barausse and Yagi~\cite{Barausse:2015wia} have shown that in
shift-symmetric Horndeski theories, under certain assumptions, the
stellar sensitivity (which quantifies the dependence of the star's
gravitational mass on the background scalar field)
vanishes. Sensitivities are important in a post-Newtonian (PN)
expansion of scalar-tensor theories, since they source the
leading-order emission of dipolar scalar radiation. Since the
sensitivities vanish, the dipolar energy flux also vanishes in this
subclass of Horndeski theory, and at leading order in the PN expansion
the dynamics of the binary in shift-symmetric Horndeski gravity will
be the same as in GR. The emission of gravitational waves only differs
from GR (if at all) at higher PN orders.
For the EdGB model considered by Sotiriou and Zhou, it was recently
shown~\cite{Yagi:2015oca} that NSs in this theory do not have scalar charge,
therefore the theory evades the constraints on the presence of dipolar radiation
in binary pulsars~\cite{Yagi:2011xp,Pani:2011xm}. These results are
in contrast with, for instance,
the popular ``spontaneous scalarization'' model by Damour and
Esposito-Far\`ese~\cite{Damour:1993hw,Damour:1996ke}, which does not
possess shift-symmetry, resulting in emission of dipolar scalar
radiation and allowing pulsar observations to put stringent
constraints on the theory~\cite{Freire:2012mg}. It will be interesting
to compute sensitivities and gravitational wave emission in Horndeski
theories without shift symmetry.

\section{Concluding Remarks}
\label{sec:conclusions}

In this paper we reviewed our current understanding of compact objects
(BHs and NSs) in Horndeski gravity. Despite of the complexity of the
Horndeski action, there has been rapid progress in specific subclasses
of the theory, but there are still many open problems.

Existing BH solutions were found under the assumption of shift and/or
reflection symmetry. It would be interesting to find BH solutions in
theories that do not satisfy these simplifying assumptions, and to
study their stability properties. More in general, studies of
stability and dynamics (including quasinormal mode calculations) are
in their infancy. It is important to understand if BH solutions in
generic Horndeski theories differ significantly from GR in terms of
their structure and dynamics, and if so, whether they could leave
observable imprints in astrophysical
settings~\cite{Berti:2015itd,Johannsen:2015mdd,Yagi:2016jml,Bambi:2015kza}.

Stellar models have also been constructed in just a few special
cases. We are currently working to extend these studies to more
general classes of theories and more realistic equations of
state~\cite{MaselliPrep}. Our main goal is to understand whether stars
in Horndeski gravity can exhibit observable deviations from GR in the
strong-field regime. Some classes of Horndeski gravity may produce
phenomena similar to spontaneous
scalarization~\cite{Damour:1993hw,Damour:1996ke}, producing observable
signatures in (say) Advanced LIGO while being compatible with
weak-field bounds: see e.g. recent work on the ``asymmetron''
scenario~\cite{Chen:2015zmx} and massive scalar-tensor theories
\cite{Alsing:2011er,Ramazanoglu:2016kul,Yazadjiev:2016pcb} for similar
proposals.

\section*{Acknowledgments}

A.M. was supported by NSF Grants No. 1205864, 1212433 and 1333360. E.B. was
supported by NSF CAREER Grant No. PHY-1055103 and
by FCT contract IF/00797/2014/CP1214/CT0012 under the IF2014 Programme. H.O.S
was supported by NSF CAREER Grant No. PHY-1055103 and by a Summer Research
Assistantship Award from the University of Mississippi. M.M. was supported
by the FCT-Portugal through Grant No. SFRH/BPD/88299/2012. H.O.S would like
to thank the organizing committee of the III Amazonian Symposium on Physics,
V NRHEP Network Meeting and the Federal University of Par\'a for the hospitality.

\end{document}